\documentclass[12pt,mathptm,times]{article}
\usepackage{graphicx}
\usepackage{setspace}
\usepackage[round,numbers]{natbib}
\usepackage{amsmath, amsthm, amsfonts, amssymb}
\usepackage{graphicx}

\newcommand{\vA}{\mathbf{A}}
\newcommand{\vD}{\mathbf{D}}
\newcommand{\vF}{\mathbf{F}}
\newcommand{\vG}{\mathbf{G}}
\newcommand{\vI}{\mathbf{I}}
\newcommand{\vM}{\mathbf{M}}
\newcommand{\vP}{\mathbf{P}}

\newcommand{\vT}{\mathbf{T}}

\newcommand{\vW}{\mathbf{W}}
\newcommand{\vu}{\mathbf{u}}
\newcommand{\vv}{\mathbf{v}}
\newcommand{\vone}{\mathbf{1}}

\newcommand{\vkappa}{\boldsymbol\kappa}

\newcommand{\trsp}{\mathsf{T}}

\DeclareMathOperator{\diag}{diag}

\title{Evolution on neutral networks accelerates the ticking rate of the molecular clock}
\author{Susanna Manrubia$^1$ and Jos\'e A. Cuesta$^{2,3}$\\
Grupo Interdisciplinar de Sistemas Complejos (GISC), Madrid\\
$^1$ Dept. de Biolog\'{\i}a de Sistemas, Centro Nacional de Biotecnolog\'{\i}a (CSIC)\\
c/ Darwin 3, 28045 Madrid, Spain. \\
$^2$ Dept. de Matem\'aticas, Universidad Carlos III de Madrid\\
28911 Legan\'es, Madrid, Spain. \\
$^3$ Instituto de Biocomputaci\'on y F\'\i sica de Sistemas Complejos (BIFI)\\ 
Universidad de Zaragoza, 50009 Zaragoza, Spain.
}
\date{\today}

\begin{document}

\maketitle

\begin{abstract}
Large sets of genotypes give rise to the same phenotype because phenotypic
expression is highly redundant. Accordingly, a population can accept mutations
without altering its phenotype, as long as the genotype mutates into
another one on the same set. By linking every pair of genotypes that are
mutually accessible through mutation, genotypes organize themselves into
neutral networks (NN). These networks are known to be heterogeneous and
assortative, and these properties affect the evolutionary dynamics of the
population. By studying the dynamics of populations on NN with arbitrary
topology we analyze the effect of assortativity, of NN (phenotype) fitness, 
and of network size. We find that the probability that the population leaves 
the network is smaller the longer the time spent on it. This progressive
``phenotypic entrapment'' entails a systematic increase in the overdispersion
of the process with time and an acceleration in the fixation rate of neutral
mutations. We also
quantify the variation of these effects with the size of the phenotype and with 
its fitness relative to that of neighbouring alternatives.
\end{abstract}


\section{Introduction}

The relationship between genotype and phenotype, the ways in which this map
conditions the adaptive dynamics of populations, or the imprints that
life-histories leave in the genomes of organisms are essential questions to be
solved before a complete evolutionary theory can be achieved. Genotypes, which
encode much of the information required to construct organisms, are
occasionally affected by mutations that modify the phenotype, their visible
expression and the target of natural selection. Many mutations are neutral
instead~\cite{kimura:1968}, varying the regions of the space of genotypes that
can be accessed by the population~\cite{nimwegen:1999} and conditioning its
evolvability~\cite{draghi:2011} but leaving the phenotype unchanged.  The
relation between genotype and phenotype is not one-to-one, but many-to-many. In
particular, genotypes encoding a specific phenotype may form vast, connected
networks that often span the whole space of possible genotypes. The existence
of these networks in the case of proteins was postulated by Maynard
Smith~\cite{maynard-smith:1970} as a requirement for evolution by natural
selection to occur. Subsequent research has shown that these networks do exist
for functional proteins~\cite{lipman:1991}, for other macromolecules as
RNA~\cite{schuster:1994}, and generically appear in simple models of the
genotype-phenotype map mimicking regulatory gene
networks~\cite{ciliberti:2007}, metabolic reaction
networks~\cite{matias-rodriguez:2009}, or the self-assembly of protein
quaternary structure~\cite{greenbury:2014}.

Nevertheless, systematic explorations of the topological structure of neutral
networks have been undertaken only recently, despite the fact that some of the
implications of neutral network structure on sequence evolution were identified
long ago. For instance, Kimura's neutral theory~\cite{kimura:1968} postulated
that the number of neutral substitutions in a given time interval was Poisson
distributed. That assumption had an underlying hypothesis that was not
explicitly stated at the time, namely, that the number of neutral mutations
available to any genotype was constant, independent of the precise genotype, of
time, or of the expressed phenotype. In other words, neutral networks were
assumed to be homogeneous in degree. A consequence was that the variance of the
number of mutations accumulated should equal the mean, and the dispersion index
($R$, the ratio between the variance and the mean) must then be equal to 1.
Very early, however, it was observed that $R$ was significantly larger than 1
in almost all cases analyzed~\cite{kimura:1971,langley:1974,gillespie:1984}.
The appearance of short bursts of rapid evolution was ascribed to
episodes of positive Darwinian selection~\cite{gillespie:1984} that may reflect
fluctuations in population size in quasi-neutral environments, where epistatic
interactions would become relevant~\cite{takahata:1987}.

The fact is that neutral networks are highly heterogeneous. Some genotypes are
brittle and easily yield a different phenotype under mutation. They have one or
few neighbours within the NN. Other genotypes, instead, are robust and can
stand a very large number of mutations while maintaining their biological or
chemical function.  
The existence of
variations in the degree of neutrality of genotypes (in their robustness) was
soon put forth as a possible explanation for the overdispersion of the
molecular clock~\cite{takahata:1987}. Nowadays, the distributions of robustness
of the genotypes in several different NN have been measured, turning out to be
remarkably broad~\cite{bastolla:2003,aguirre:2009,wagner:2011}. The effects of
fluctuating neutral spaces in the overdispersion of the molecular clock have
been investigated in realistic models of evolution for
proteins~\cite{bastolla:1999} and quasispecies~\cite{wilke:2004}, and also from
a theoretical viewpoint~\cite{raval:2007}. 

In this contribution, we explore three features of neutral networks whose
consequences in the rate of fixation of mutations have not been systematically
investigated. They are (i) the correlations in neutrality between neighbouring
genotypes, (ii) the degree of redundancy of phenotypes (the size of NN), and
(iii) the fitness of the current phenotype in relation to accesible
alternatives.  First, the degree of neutrality is not randomly distributed in 
NN. Thermodynamical arguments~\cite{bornberg-bauer:1999,wuchty:1999} and 
analyses of full NN~\cite{aguirre:2011} indicate that genotypes tend to cluster
as a function of their robustness, implying that NN belong to the class of the 
so called assortative networks.  It is known that populations evolving on 
neutral networks tend to occupy maximally connected regions in order to minimize
the number of mutations changing their phenotype~\cite{nimwegen:1999}. In
assortative networks, neutral drift entails a canalization towards 
mutation-selection equilibrium, progressively increasing the rate of fixation 
of neutral mutations through a dynamical process that we dub {\it phenotypic 
entrapment}. Second, there is abundant evidence coming from computational 
genotype-phenotype maps~\cite{li:1996,bloom:2007,aguirre:2011,greenbury:2014} 
and from empirical reconstruction of neutral networks~\cite{dallolio:2014} that 
the average robustness of a given phenotype grows with the size of its 
associated NN: here we quantify the
effect of a systematic difference in average degree on the probability of
fixation of neutral mutations.  Third, the difference in fitness between
genotypes in the current NN and their mutational neighbours affects the
probability that a mutation (be it neutral, beneficial, or deleterious) gets 
fixed in the population, and with it, as we explicitly show, the rate of the
molecular clock during intervals of strictly neutral drift.

With the former goals in mind, we develop an out-of-equilibrium formal
framework to describe the dynamics of homogeneous, infinite populations 
on generic NN. We demonstrate that the population keeps memory of its past history 
since, as time elapses, the likelihood that it visits genotypes of increasingly higher
robustness augments in a precise way that we calculate. This is a consequence
of assortativity and a dynamic manifestation of the ``friendship
paradox''~\cite{feld:1991} described in social networks (your friends have more
friends than you). As a result, the probability that the population leaves
the network explicitly depends on the elapsed time. Further, the decline of
this probability with time entails a systematic acceleration of the rate of
accumulation of neutral mutations. The degree of entrapment is higher the
larger the neutral network and the broader the difference between the fitness
of the current phenotype and that of accessible alternatives. These results are
fairly general and have implications in the derivation of effective models of
phenotypic change and in the calibration of molecular clocks.

\section{Dynamical model}

\subsection{Description of neutral networks and populations}

In the forthcoming sections, we will employ terms related to dynamics on
complex networks to describe the dynamical process undergone by a homogeneous
population on a NN. Genotypes are the nodes of the network, and two nodes are
linked if the corresponding genotypes can be mutually accessed through a single
mutation. Links are all identical, so we assume that the mutational process
affects all genotypes in the same way. The robustness of a node is a quantity
proportional to its degree $k$, that is, to the number of links leading to
other nodes within the NN (or the number of neutral mutations). 

For the sake of analytical tractability, we will use a mean-field-like
description of the NN. This amounts to assuming that all nodes of the same
degree (robustness) are equivalent with respect to the dynamics, and therefore
the network can be characterized only through its degree distribution, $p(k)$
(the probability that a node has degree $k$, or the distribution of robustness
values), and its nearest neighbour degree distribution $p_{\text{nn}}(k|k')$
(the probability that a degree $k'$ node's neighbour has degree $k$), where
$k,k'=1,\dots,z$. We are assuming that there is a maximum degree $z$ in the NN.
This is always true if genotypes are sequences whose loci are taken from a finite
alphabet (e.g. $\{0,1\}$, nucleotides or aminoacids). If there are $A$
different types in the alphabet, sequences have length $L$, and mutations refer
just to point mutations (the most widely studied mutational process), then $z =
(A-1) L$. Typically, thus, $z\gg 1$.   

The strength of correlation between the degree of neighbouring nodes is
measured through the {\it degree-degree correlation coefficient} $r$. This is a
quantity that characterizes the assortativity of a network and is defined as
the Pearson correlation coefficient (c.f. the covariance) between the degree of
nodes and the degree of their neighbours. For a network with $i=1, \dots, M$
nodes, each with degree $k_i$ and whose neighbours have on average degree
$k_i^{nn}$, $r$ is defined as
\begin{equation}
r = \frac{\sum_{i=1}^M\left(k_i - \overline{k}\right)\left(k_i^{nn} -
\overline{k^{nn}}\right)} {\sqrt{\sum_{i=1}^M\left(k_i - \overline{k}\right)^2}
\sqrt{\sum_{i=1}^M\left(k_i^{nn} - \overline{k^{nn}}\right)^2}},
\end{equation}
with $\overline{k}=M^{-1}\sum_i k_i$, $\overline{k^{nn}}= M^{-1}\sum_i
k_i^{nn}$.  The largest value of assortativity $r=1$ is achieved when the
average degree of neighbours of node $i$ equals its degree, $k_i^{nn}=k_i$.
Anticorrelations in degree yield negative values of $r$. By definition, $-1 \le
r \le 1$.

We assume that all genotypes in a NN have the same fitness, so evolution is 
strictly neutral within the network. We should only take it into account 
non-neutral mutations that get fixed, that is, that move the population from a 
node in its current NN to a
node of a different NN. As of neutral evolution, population dynamics
distinguishes two basic regimes. If $\mu$ denotes the mutation rate and $N_e$
the (effective~\cite{ewens:2004}) population size, those correspond to low
($\mu N_e \ll 1$) and high ($\mu N_e \gg 1$) mutation rates. For high mutation
rate populations are heterogeneous and we need an ensemble of individuals
coupled through replication and mutation to account for their evolution. This
is a complex situation where quasispecies dynamics comes into play, and one has
to resort to computational simulations to describe its
dynamics~\cite{wilke:2004}. In this study, we will work in the limit $\mu N_e
\ll 1$, where every mutation either gets fixed in or disappears from the
population before a new mutation occurs, so the population is always
homogeneous and its evolution is effectively described as a single {\it random
walk} (RW) on the NN.

\subsection{Dynamics}

We will describe the dynamical process that begins with the arrival of a random
walker to the NN and ends when it leaves to a neighbouring NN. In particular,
we aim at describing the entrapment effect that occurs while the walker remains
within the NN. Let us mention at this point a fact that will become clear later
on: because of frequent jumps in and out of NNs, populations will hardly spend
a time long enough within a single NN to reach the equilibrium distribution.
Therefore, the relevant process to describe is the transient dynamics that
starts with the arrival to the NN, instead of the asymptotic behaviour within
the phenotype. For this reason, we require a good estimation of the initial
probability distribution of arrival at any node of the NN.

The process starts as soon as the walker jumps to a node of the NN from
outside. We will assume that the initial jump is equally likely to occur through 
any link of the NN pointing outwards, so the probability that a node of degree $k$
is chosen as the entry node should be proportional to the number of outbound
links $z-k$ and to the abundance of nodes of degree $k$.  This yields the
initial condition
\begin{equation}
P_0(0)=0, \quad P_k(0)=\frac{z-k}{z-\langle k\rangle}p(k),
\quad k=1,\dots,z.
\label{eq:P0}
\end{equation}
The state $i=0$ does not describe any node, but it will prove convenient to
introduce it to represent the exterior of the NN. Hence $P_0(t)$ will be the
probability that the process ends at any time $t>0$; the choice $P_0(0)=0$
means that the process begins as soon as it ``leaves'' the exterior.

Since the mean field description assumes that all nodes of the same degree are
equivalent, the RW on the NN can be further described as a stochastic process
$X(t)$ taking values in the set $\{0,1,\dots,z\}$. These values represent the
degree of the node the walker is in at time $t$,  with state $0$ corresponding
to the exterior of the NN, as just described. When the walker jumps to state
$0$ the phenotype changes, and other considerations are needed to follow up the
evolution. For our purposes the process will end at this point, so $0$ will be
an {\it absorbing state} (the only one in this process).

The stochastic process just introduced is a continuous-time Markov chain with
transition rate
\begin{equation}
\Pr\{X(t+\delta t)=k|X(t)=k'\}=\mu_kW_{kk'}\delta t+o(\delta t) \, .
\end{equation}
Element $ W_{kk'}$ of matrix $\vW$ is the conditional probability that, given
that a jump occurs when the process is at a node of degree $k'$, it ends up at a
node of degree $k$. Within our mean-field description, this probability can be
simply obtained as
\begin{equation}
W_{kk'}= \frac{k'}{z}p_{\text{nn}}(k|k'), \quad k,k'>0, \quad
W_{0k'}= 1-\frac{k'}{z}, \quad k'>0, \quad
W_{00}=0.
\label{eq:W}
\end{equation}
As above, the rationale for this choice is that $z-k'$ links of the node where
the RW is located point outside the NN and $k'$ point inside, and if the
process stays within the NN (probability $k'/z$), the probability that the node
it jumps to has degree $k$ is $p_{\text{nn}}(k|k')$.  The nearest-neighbour
degre distribution $p_{\text{nn}}(k|k')$ is zero if $p(k')=0$ and is normalized
otherwise; therefore
\begin{equation}
\sum_{k=0}^zW_{kk'}=1, \text{ for all $k'$ such that } p(k')>0.
\end{equation}

\subsection{Leaving the NN}

Let us now discuss the meaning of the transition rates $\mu_k$. If $k>0$ the RW
jumps to a node that belongs to the NN and thus the mutation is neutral. Hence,
$\mu_k=\mu$ because in that case the mutation rate is also the rate of fixation
of neutral mutations in a population of arbitrary size~\cite{gillespie:1994}.
On the other hand, $\mu_0$ implies a jump outside the NN, hence a phenotypic
change. The new phenotype will have in general a different fitness, and the
rate at which non-neutral mutations go to fixation in a population of effective
size $N_e$ is $\mu_0=\mu N_e(f-1)/(f^{N_e}-1)$, where $f$ is the fitness of the
current phenotype relative to that of the phenotype the process jumps
to~\cite{ewens:2004}. The ratio $\phi\equiv\mu_0/\mu$ can take any value in the
interval $0\leqslant\phi\leqslant N_e$. The lower bound occurs if purifying
selection eliminates any phenotype other than the current one; the upper bound
is the rate at which the population adopts a highly fitter new phenotype. Neutral
changes of phenotype correspond to $\phi=1$.

There is a simplification implicit in the expression of $\phi$, though. If the
neighbouring NNs are diverse in fitness values, $\phi$ should be replaced by an
appropriate average over this diversity---weighted with the probabilities of
jumping to these other NNs. In any case some factor $0<\phi<N_e$ will account
for the differences in fitness between the current NN and those of its
neighbouring NNs. For the discussion to come we do not need to be more specific
about its precise form.

\subsection{Master equation}

The probability distribution of the process is $P_k(t)\equiv\Pr\{X(t)=k\}$,
which stands for the probability that at time $t$ the process sits on a node of
degree $k=1,\dots,z$, or is outside the NN, if $k=0$. This probability
satisfies the master equation
\begin{equation}
\dot P_k=\sum_{j=0}^z[\mu_kW_{kj}P_j-\mu_jW_{jk}P_k], \quad
k=0,1,\dots,z.
\label{eq:master}
\end{equation}

A few considerations transform \eqref{eq:master} into a handier expression for
further calculations. To begin with, we can set $\mu=1$, i.e., measure time in
units of $\mu^{-1}$. Let us also introduce the \emph{support} of $p(k)$, i.e.,
the set $\mathcal{S}\equiv\{k\,:\,1\leqslant k\leqslant z\,, \,p(k)>0\}$. Thus,
any $k\notin\mathcal{S}$ has $p(k)=0$, $p_{\text{nn}}(j|k)=0$, and, given
definition \eqref{eq:P0}, $P_k(0)=0$.  Therefore, according to \eqref{eq:W},
for any $k\notin\mathcal{S}$ the master equation \eqref{eq:master} becomes
$\dot P_k=-[\phi+(1-\phi)k/z]P_k$, whose solution is
$P_k(t)=P_k(0)e^{-[\phi+(1-\phi)k/z]t}=0$.

Because of this result we can ignore in \eqref{eq:master} any index
$k\notin\mathcal{S}$, that is, those degrees that are not represented by any
node in the NN. Accordingly, if we introduce vector $\vP\equiv(P_k)^{\trsp}$,
$k\in\mathcal{S}$, and matrices
\begin{equation}
\vF=\diag\left(\phi+(1-\phi)\frac{k}{z}\right), \qquad
\vA=\left(p_{\text{nn}}(k|j)\frac{j}{z}\right), \quad
k,j\in\mathcal{S},
\end{equation}
the master equation \eqref{eq:master} simply becomes
\begin{equation}
\dot{\vP}=-(\vF-\vA)\vP.
\label{eq:mastermat}
\end{equation}

\section{Results}

\subsection{Time spent on the NN}

An important quantity characterizing the process is
the probability that, after a time $t$, the population is still on the NN.
This probability can be calculated as $Q(t)\equiv 1-P_0(t)= \vone\cdot\vP(t)$,
where $\vone=(1,\dots,1)^{\trsp}$. For $k=0$, eq.~\eqref{eq:master} yields
$\dot Q=\phi[\vkappa\cdot\vP(t)-Q]$, where $\vkappa=(k/z)^{\trsp}$,
$k\in\mathcal{S}$, a differential equation which can be rewritten as the
integral equation
\begin{equation}
Q(t)=e^{-\phi t}+\phi\int_0^te^{-\phi(t-\tau)}\vkappa\cdot\vP(\tau)\,d\tau.
\label{eq:Qgeneral}
\end{equation}

In appendix~\ref{app:A} we prove that the eigenvalues $\alpha_k$ of matrix
$\vF-\vA$ are all real and different. Therefore, denoting $\vu_k$ and $\vv_k$
the corresponding right and left eigenvectors, respectively, we can write
\begin{equation}
Q(t)=\sum_{k\in\mathcal{S}}q_ke^{-\alpha_k t}, \qquad
q_k\equiv\frac{[\vv_k\cdot\vP(0)](\vu_k\cdot\vone)}{(\vv_k\cdot\vu_k)},
\quad k\in\mathcal{S}.
\label{eq:Q}
\end{equation}
(Notice that $\sum_{k\in\mathcal{S}}q_k=Q(0)=1$.)

Equation~\eqref{eq:Q} makes apparent that, in general, the life-time
distribution of the process before leaving the NN is not exponential, but
displays many different time scales, as many as elements in the set
$\{\alpha_k\}$. Asymptotically, $Q(t)\sim q_{\text{min}}
e^{-\alpha_{\text{min}}t}$, with $\alpha_{\text{min}}$ the smallest eigenvalue.
Thus, the rate at which the process leaves the NN tends to a minimum as time
elapses---in other words, it gets more and more \emph{trapped} within the NN.
Eventually, mutation-selection equilibrium is reached and the process can be
described through a unique time scale, $\alpha_{\rm min}$. 

\subsection{Mutation accumulation}

Perhaps the most immediate consequence of this entrapment of the evolutionary
process in a NN is that the observed rate at which neutral mutations accumulate
increases with time. Together with the stochastic nature of the dynamics, the
consequence is two-fold: on the one hand, the molecular clock gets accelerated
during those periods of strictly neutral evolution;
on the other hand, overdispersion of the molecular clock varies
non-monotonically with time, as we show in the following.

In order to analyze how mutations accumulate with time, we need to introduce
$P_k(m,t)$, the probability that the process is at time $t$ at a node of degree
$k$ of the NN, having undergone exactly $m$ mutations. Introducing
$\vP_m(t)\equiv\big(P_k(m,t)\big)^{\trsp}$, $k\in\mathcal{S}$, the dynamic
equation for this vector can be written as
\begin{equation}
\dot\vP_m= \vA\vP_{m-1}-\vF\vP_m,
\label{eq:mutations}
\end{equation}
which is valid for all integers $m$ if we assume $\vP_m(t)=0$ for all $m<0$.
Obviously $\vP_m(0)=\vP(0)\delta_{m,0}$, with $\vP(0)$ the initial probability
distribution \eqref{eq:P0}.

In order to extract information on the mean and variance of the number of
mutations accumulated at time $t$, it is useful to introduce the moment
generating function
\begin{equation}
\vG(\theta,t)\equiv \sum_m\vP_m(t)e^{m\theta},
\label{eq:generating}
\end{equation}
so that
\begin{equation}
m_r(t)\equiv \langle m^r\rangle(t)= \lim_{\theta\to 0}
\frac{\partial^r}{\partial\theta^r}
\left(\sum_{k\in\mathcal{S}} G_k(\theta,t)\right).
\label{eq:moments}
\end{equation}

Multiplying eq.~\eqref{eq:mutations} by $e^{m\theta}$ and summing on $m$ we
obtain the dynamic equation for $\vG(\theta,t)$
\begin{equation}
\frac{\partial}{\partial t}\vG(\theta,t)=\left(e^{\theta}\vA-\vF\right)
\vG(\theta,t).
\label{eq:G}
\end{equation}
Setting $\theta=0$ shows that $\vG(0,t)=\vP(t)$, the solution of
\eqref{eq:mastermat}, hence $m_0(t)=Q(t)$.

Differentiating \eqref{eq:G} with respect to $\theta$ yields, for $\theta=0$,
\begin{equation}
\begin{split}
\frac{\partial}{\partial t}\vG_{\theta}(0,t) &=
(\vA-\vF)\vG_{\theta}(0,t)+\vA\vG(0,t),  \\
\frac{\partial}{\partial t}\vG_{\theta\theta}(0,t) &=
(\vA-\vF)\vG_{\theta\theta}(0,t)+\vA[2\vG_{\theta}(0,t)+\vG(0,t)]
\end{split}
\label{eq:Gthth}
\end{equation}
(subscripts $\theta$ denote partial derivatives).

Simultaneously solving eqs.~\eqref{eq:mastermat}, and \eqref{eq:Gthth}, and
using \eqref{eq:moments} yields $m_0(t)=Q(t)$, $m_1(t)$ and $m_2(t)$. We are
not yet done, since these are not the quantities required to estimate
magnitudes related to the molecular clock: Actual measurements are always
performed on extant populations, so we need the mean $m(t)$ and variance $v(t)$ in
the number of mutations conditioned on remaining within the NN at time $t$.
These are obtained as
\begin{equation}
m(t)=\frac{m_1(t)}{m_0(t)}, \qquad
v(t)=\frac{m_2(t)}{m_0(t)}-m(t)^2.
\end{equation}
Correspondingly, the overdispersion of the molecular clock at time $t$ should
be computed as
\begin{equation}
R(t)\equiv \frac{v(t)}{m(t)}=\frac{m_2(t)}{m_1(t)}-\frac{m_1(t)}{m_0(t)}.
\label{eq:overdispersion}
\end{equation}

\section{Examples}

In this section we will compare analytical and numerical results for the
dynamics of a RW (representing a homogeneous population) on networks differing
in size, fitness, and topological features. Our aim is to better understand the
quantitative effect of these variables on the dynamics of the population,
paying particular attention to the time scales involved, to the change in
overdispersion with time, and to variations in the rate of fixation of mutations. 
We will analyse a random network (with near zero
assortativity and a well defined average degree) as well as two highly
assortative networks with either only two degrees or a constant degree
distribution. Our final example comes from secondary structure RNA networks,
which share several properties with other natural NN. 

\subsection{Random network}

Despite its apparent simplicity, Eq.~(\ref{eq:mastermat}) can not be
analytically solved in general. An interesting exception is the case of a
random network constructed by randomly drawing links between pairs of nodes. In
this situation, the degree follows a Poisson distribution and, consequently,
the average degree is well defined. Even this special case is exactly solvable
only for $\phi=1$. 

In a random network there is no correlation between the degree of a given node
and that of its nearest neighbours. If the network is infinitely large, its
assortativity $r=0$---though a residual assortativity will be typically
obtained when the number of nodes is finite. This lack of correlation reflects
in that the probability that a node's neigbour has degree $k$ only depends on
its own degree. As a matter of fact, $p_{\text{nn}}(k|j) =(k/\langle k\rangle)
p(k)$. (Here, and in what follows, $\langle k^r\rangle\equiv\sum_kk^rp(k)$.)
For this special case,
\begin{equation}
\vF=\vI, \qquad \vA=\left(\frac{kj}{z\langle k\rangle}p(k)\right),
\end{equation}
and deriving an expression for $Q(t)$ is straightforward (see
Appendix~\ref{app:B}). The resulting formula is
\begin{equation}
Q(t) = \frac{z\rho-\langle k\rangle}{\rho(z-\langle k\rangle)} e^{-t}
+ \frac{(1-\rho)\langle k\rangle}{\rho(z-\langle k\rangle)}e^{-(1-\rho)t},
\qquad
\rho\equiv\frac{\langle k^2\rangle}{z\langle k\rangle}.
\label{eq:QtER}
\end{equation}

Despite the ease of this case, and the fact that the NN has no fitness
advantage with respect to the average neighbouring NN ($\phi=1$), we can
already observe what turns out to be a generic property of heterogeneous
networks: the lifetime of the RW within the NN is not exponentially
distributed. For random networks, there are only two time scales: the
``standard'' one, $\mu^{-1}$, and a slower one, with characteristic time
$(1-\rho)^{-1}\mu^{-1}$. The reason for this non-exponential behaviour---and
for the equivalent effect that will arise in any heterogeneous network---is
that the RW comes in the NN preferentially through nodes of low inner connectivity 
(high outer
connectivity) and as time passes without leaving the NN it progressively moves
toward the regions of higher connectivity (notice that the mean degree of a
node's neighbour is $\langle k^2\rangle/\langle k\rangle>\langle k\rangle$),
from which escaping is less probable. Thus, the first scale corresponds to
walkers which leave the network in very few steps, whereas the second (slow)
scale characterizes walkers that have spent some time in the NN.

Further insight can be gained by calculating the overdispersion
\eqref{eq:overdispersion} (see Appendix~\ref{app:B}), whose expression for this
case becomes
\begin{equation}
R(t)=1+\frac{(z\rho-\langle k\rangle)\rho te^{-\rho t}}
{(1-\rho)\langle k\rangle+(z\rho-\langle k\rangle)\rho te^{-\rho t}}.
\label{eq:overdispersionrandom}
\end{equation}

\begin{figure}[ht]
\centerline{\includegraphics[width=70mm]{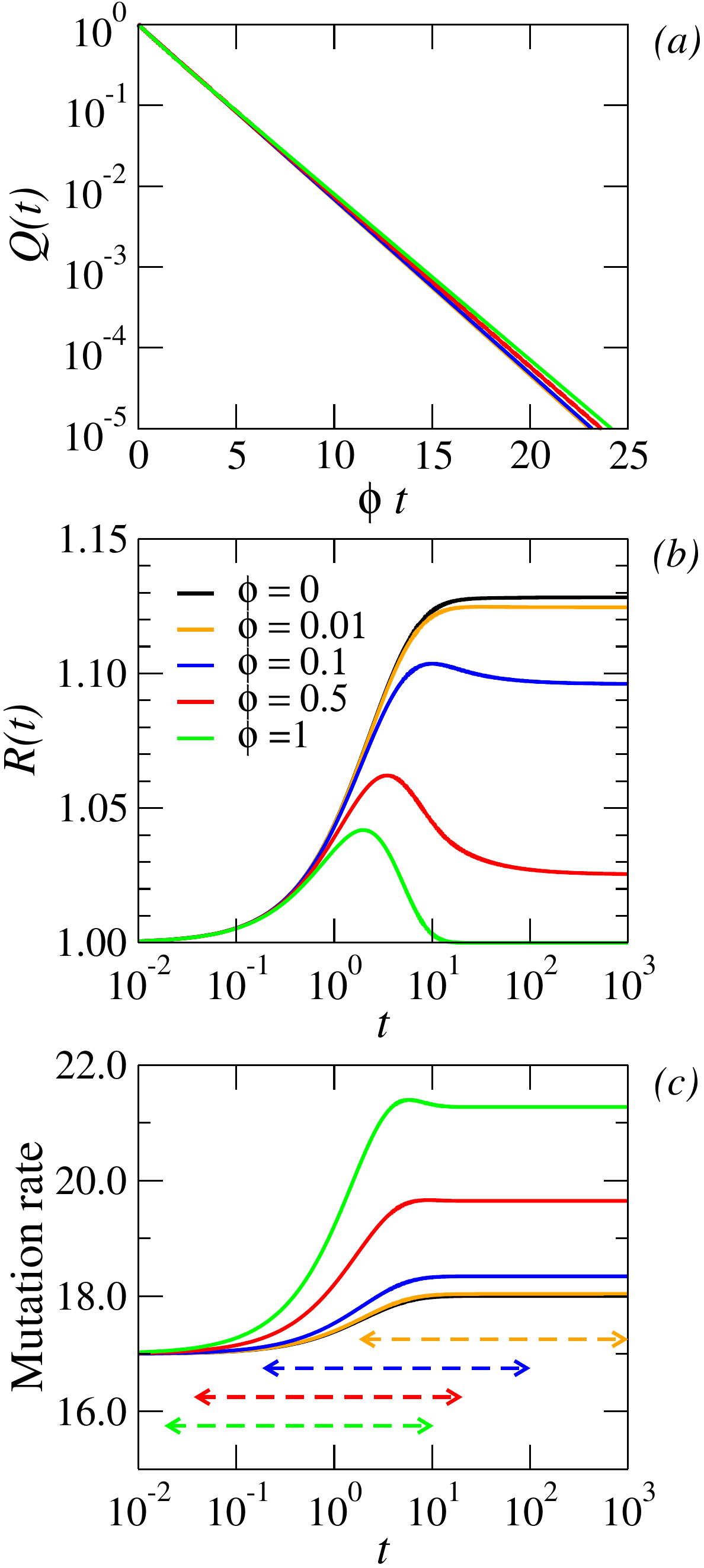}}
\caption{Dynamics of RWs on an Erd\H{o}s-Renyi random network with $M=20 000$
nodes and average degree $\langle k\rangle=18$. (a) Probability $Q(t)$ that the
RW stays on the network after a time $t$. (b) Overdispersion of the network.
(c) Mutation rate or acceleration of the molecular clock. The time 
interval through
which the number of trajectories remaining on the network decreases from $99\%$
to $1\%$ is represented as a horizontal dashed line limited by two arrows, with the
same colour as the $\phi$ curve they correspond to. The same code is used in
subsequent corresponding panels for other networks.
The measured assortativity for this finite network is $r\approx 0.018$.}
\label{fig:ERnetwork}
\end{figure}

Figure~\ref{fig:ERnetwork} shows the probability $Q(t)$ to remain on the
network at time $t$, the overdispersion $R(t)$ of the molecular clock with
time, and the mutation rate (measured as the time derivative of the mean
number of mutations) for an Erd\"os-Renyi random network. In
Fig.~\ref{fig:ERnetwork}(a), as well as in all other plots, $Q(t)$ is
represented as a function of the rescaled variable $\phi t$. The reason is
that, as can be seen, there is a partial collapse of all $Q(t)$ curves (but see
the discussion for forthcoming examples). Also, the case $\phi=0$ is not
included, since $\phi=0$ implies that the RW cannot leave the NN and,
therefore, $Q(t)=1$, for all $t$. For Erd\"os-Renyi random networks, the fast
time scale is barely visible, due to the fact that the RW becomes rapidly
trapped. The slow time scale actually dominates the dynamics: for the values
chosen, $(1-\rho) \simeq 0.47$ (case $\phi=1$) and the coefficient of the
corresponding exponential in $Q(t)$ is about ten times larger than that of the
fast time scale.

Figure~\ref{fig:ERnetwork}(b) shows the typical behavior of
the overdispersion on heterogeneous NN: an initial increase up to a maximum
followed by a decay to an asymptotic value. Interestingly, even for $\phi=1$
(corresponding to Eq.~\eqref{eq:overdispersionrandom}) a $4\%$ overdispersion
is reached before it decays to its asymptotic value, a process that requires
about ten mutations ($t\sim 10\mu^{-1}$). As values of $\phi$ diminish, the
asymptotic value of $R(t)$ monotonically increases.
Three features that will show up in the remaining examples as well should be
highlighted: firstly, $\phi$ has a strong, nontrivial effect on overdispersion; 
secondly, it is in the transient where overdispersion reaches its largest value;
thirdly, the rate of accumulation of mutations increases with $\phi$ (it therefore 
diminishes the larger the difference between the fitness of the current NN and 
its neighbouring phenotypes).

In Figure~\ref{fig:ERnetwork}(c) we can appreciate how the ticking rate of the
molecular clock undergoes an acceleration from its initial value, eventually 
reaching a steady rate higher than 
the one it started off from. The change occurs at a time $t \sim 10\mu^{-1}$,
which corresponds to the number of mutations accumulated. Notice that, due to the
heterogeneity of the network, the initial mutation rate is already higher than 
$\mu$. As we stated earlier, the stationary state is not attained in 
general. Now we can quantify this statement by estimating the time 
$t_{1\%}(\phi)$ by 
which only 1\% of the RWs remain on the network. This time can be obtained by
numerically solving $Q(t_{1\%})=0.01$---it can be approximately read from
Figure~\ref{fig:ERnetwork}(a). In order to estimate the range of $t$
values where typical trajectories on the NN should be found, we have calculated as
well the time $t_{99\%}(\phi)$, when only $1\%$ of the trajectories have left the 
network. The interval $(t_{99\%}(\phi),t_{1\%}(\phi))$ is represented as a dashed line
within two arrows in Figure~\ref{fig:ERnetwork}(c) (and all subsequent (c) panels).
As can be seen, the maximum variation in $R(t)$ and in the mutation rate is achieved
within that time interval, while mutation-selection equilibrium is rarely attained.
Note that $t_{99\%}(\phi)$ and $t_{1\%}(\phi)$ decrease with increasing $\phi$.
The acceleration of the 
molecular clock along neutral evolution, which is the higher the smaller $\phi$, 
is a feature that will reappear in all subsequent examples, and probably stands as 
the best illustration of the phenotypic entrapment produced by NN. Typically, most 
RW will leave the network before reaching the asymptotic neutral ticking rate,
and certainly before the RW has had time to explore a significant
fraction of the NN~\cite{szollosi:2008}.

\subsection{Two-degree network}

Consider now a network with $M$ nodes such that $p(k)=1/2$ for $k=k_1,k_2$
($k_1<k_2$), and $0$ otherwise. The $M/2$ nodes of each degree are randomly
connected between them, except for $L$ links that connect (also randomly) nodes
of the two different degrees. Accordingly, $p_{\text{nn}}(k_i|k_i)=1-2L/(Mk_i)$
and $p_{\text{nn}}(k_j|k_i)=2L/(Mk_i)$. The nearest-neighbour degree
distribution clearly reflects in this case the trapping power of more connected
regions: note that the probability of entering the group of high connectivity
($k_2$) from that of low connectivity is higher than the probability of leaving
it, $p_{\text{nn}}(k_1|k_2)<p_{\text{nn}}(k_2|k_1)$. The asymmetry of
$p_{\text{nn}}$ reflects the entrapment by progressively more connected regions
of the NN. 

In terms of the variables $\ell\equiv 2L/(Mz)$, $\kappa=(k_2+k_1)/2z$ and
$\delta=(k_2-k_1)/(2z)$, matrix $\vA-\vF$ becomes
\begin{equation}
\vA-\vF=
\begin{pmatrix}
\phi(\kappa-\delta)-\ell-\phi & \ell \\
\ell & \phi(\kappa+\delta)-\ell-\phi 
\end{pmatrix},
\end{equation}
and the resulting probability \eqref{eq:Q} of remaining trapped within the
NN at time $t$ is
\begin{equation}
Q(t) = q_+e^{-\alpha_+ t}
+
q_-e^{-\alpha_- t},
\end{equation}
where
\begin{equation}
q_{\pm}=\frac{1}{2}\left[1\pm\frac{\ell-(1-\kappa)^{-1}\phi\delta^2}{\Delta}
\right], \qquad
\alpha_{\pm}=\ell+\phi(1-\kappa)\mp\Delta, 
\end{equation}
with $\Delta\equiv\sqrt{\ell^2+\phi^2\delta^2}$. Thus, the dynamics on a
two-degree network is characterized by two time scales, with coefficients
$q_{\pm}$ further determining the relative weight of each one in the process
between arriving at the NN and achieving maximum entrapment.

\begin{figure}[ht]
\centerline{\includegraphics[width=70mm]{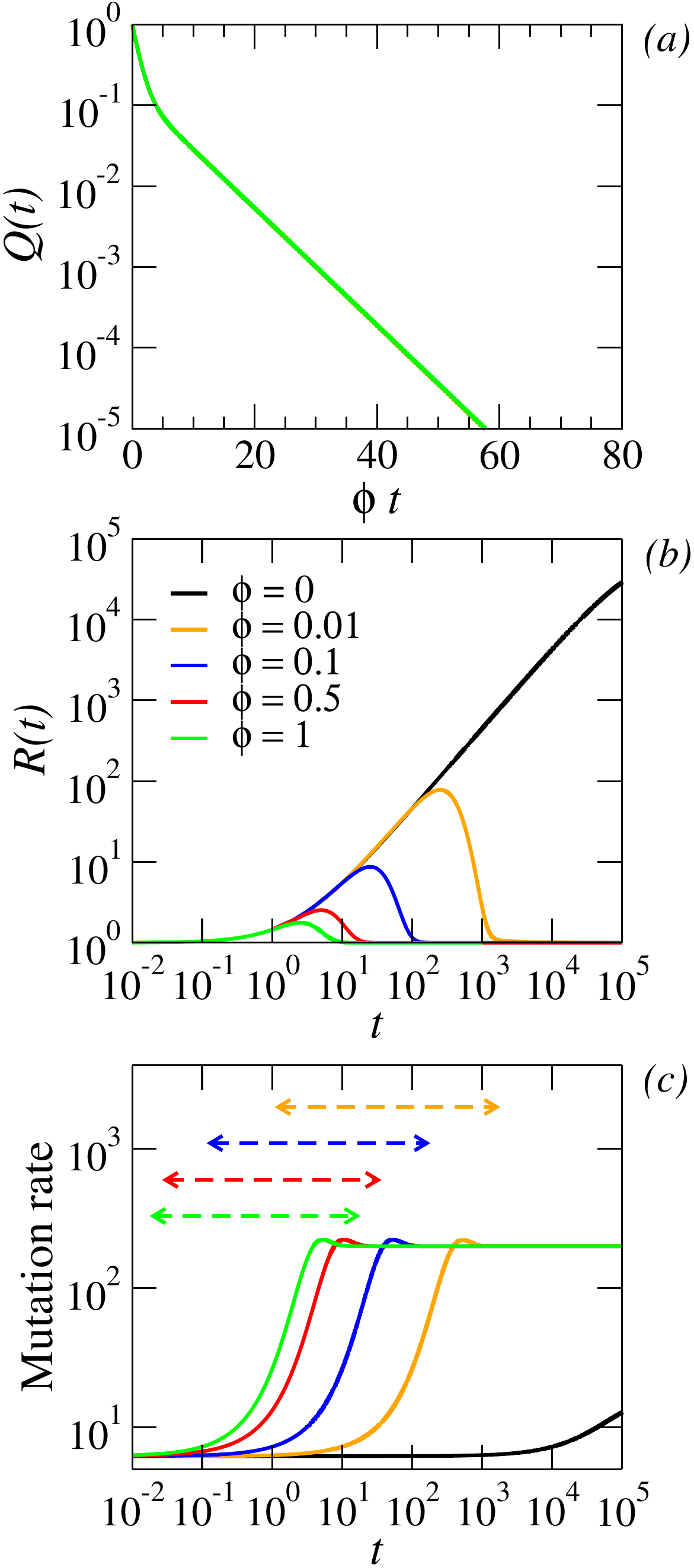}}
\caption{Dynamics of a RW on a two-degree network with $k_1=2$, $k_2=30$, 
$L=2$, and $M=20 000$ nodes. (a) Probability $Q(t)$ that the RW stays on the 
network after a time $t$. (b) Overdispersion of the network. By construction, 
the assortativity is very high, $r\approx 0.99995$.}
\label{fig:2degnetwork}
\end{figure}

Despite the formal similarity of $Q(t)$ with the solution for an Erd\"os-Renyi
network, there are important quantitative differences between the two cases.
Figure~\ref{fig:2degnetwork}(a) illustrates the presence of the two time
scales, here clearly visible (note the change in the horizontal scale with
respect to Fig.~\ref{fig:ERnetwork}(a)). The probability to stay on this
two-degree network is also significantly higher after a fixed time, which
speaks for longer trajectories in comparison to those on a random network. The
average degree does not play an essential role in this behaviour, since it has
value 16, slightly below the random network previously studied; this
nonetheless, the two-degree network is more efficient at trapping the RW. Let
us now explain why the curves for $Q(t)$ collapse so nicely in this case. This
is due to the very low value of $\ell \simeq 5 \times 10^{-6}$ for the
parameters chosen. As a result, the two time scales are well approximated as
$\alpha_{-} \simeq \phi (1-k_2/z)$ and $\alpha_+ \simeq \phi (1-k_1/z)$, thus
nearly scaling with $\phi$ as long as $\phi \gtrsim \ell$.  Hence, the collapse
of the $Q(t)$ curves is a particular feature of this network and does not hold
in general, as the next example will show. 

Differences in overdispersion between the random and the two-degree network are
still more remarkable. Figure~\ref{fig:2degnetwork}(b) (note the logarithmic
scale in the $y-$axis) represents $R(t)$, showing an extraordinary increase in
overdispersion with time, especially as $\phi$ decreases. Though not shown in
the picture, $R(t)$ eventually converges to 1, albeit the time required is
huge. This is another important difference with the random network, where the
overdispersion saturates at values $R(t \to \infty) > 1$. This is due to the
persistence of heterogeneity in the degree {\it from the viewpoint of the RW}.
Whereas in the random network the walker visits nodes of different degree once
mutation-selection equilibrium has been reached, in the two-degree network the
region of low connectivity is mostly invisible to the dynamics, trapped in the
large cluster of high-connectivity nodes. This nonetheless, there is an
enormous dispersion in the time required for a particular trajectory to become
trapped, this fact explaining the growth of overdispersion, the delayed
achievement of its maximum value, and the slow convergence to equilibrium. 

Similar differences can be observed in the mutation rate, depicted in
Figure~\ref{fig:2degnetwork}(c), compared to that for the Erd\"os-Renyi
network. The acceleration of the molecular clock for this network is very large
(over an order of magnitude with respect to the initial ticking rate). Also, as
$\phi$ decreases, the acceleration occurs at earlier times, an effect that 
was much milder in the Erd\"os-Renyi network. As a matter of fact, these curves
collapse nicely when they are represented as a funcion of $\phi t$, as it happened
with $Q(t)$. The asymptotic mutation rate is independent of $\phi$ because, similarly
to what happens with overdispersion, the RW does not see any difference in the
degrees visited at large times, even for different values of $\phi$. The next 
example clarifies how $\phi$ affects the trapping strength of nodes with different 
degree, and the resulting variation in the asymptotic mutation rate. 

\subsection{Constant degree distribution network}

This example shares some similarities to the previous one, but now nodes of all
degrees between a minimum $k_{\rm min}=2$ and a maximum value appear in equal
amounts. Also, the fraction of links connecting nodes of different degrees is
notably larger. Consider a network with a degree distribution $p(k)=1/50, 2
\le k \le 51$, $N=20,000$ nodes, and $z=52$. In order to generate an
assortative network, we proceed as follows. First, we assign an arbitrary index
to the nodes (with no meaning whatsoever) and divide them up into blocks of
400($=20000/50$) consecutive nodes. We then assign degree 2 to the first block,
degree 3 to the second block, and so on until exhausting the nodes. Now we
randomly connect the nodes, in such a way that the probability to connect a
pair of nodes at positions $i$ and $j$ is taken proportional to
$\exp\{-d/200\}$, where $d=|i-j|$. With this procedure, nodes with similar
degree are preferentially connected, and high assortativity is ensured. It is
not possible to derive any exact expression for $Q(t)$ or $R(t)$, though the
results can be interpreted in the light of our previous examples. 

\begin{figure}[ht]
\centerline{\includegraphics[width=70mm]{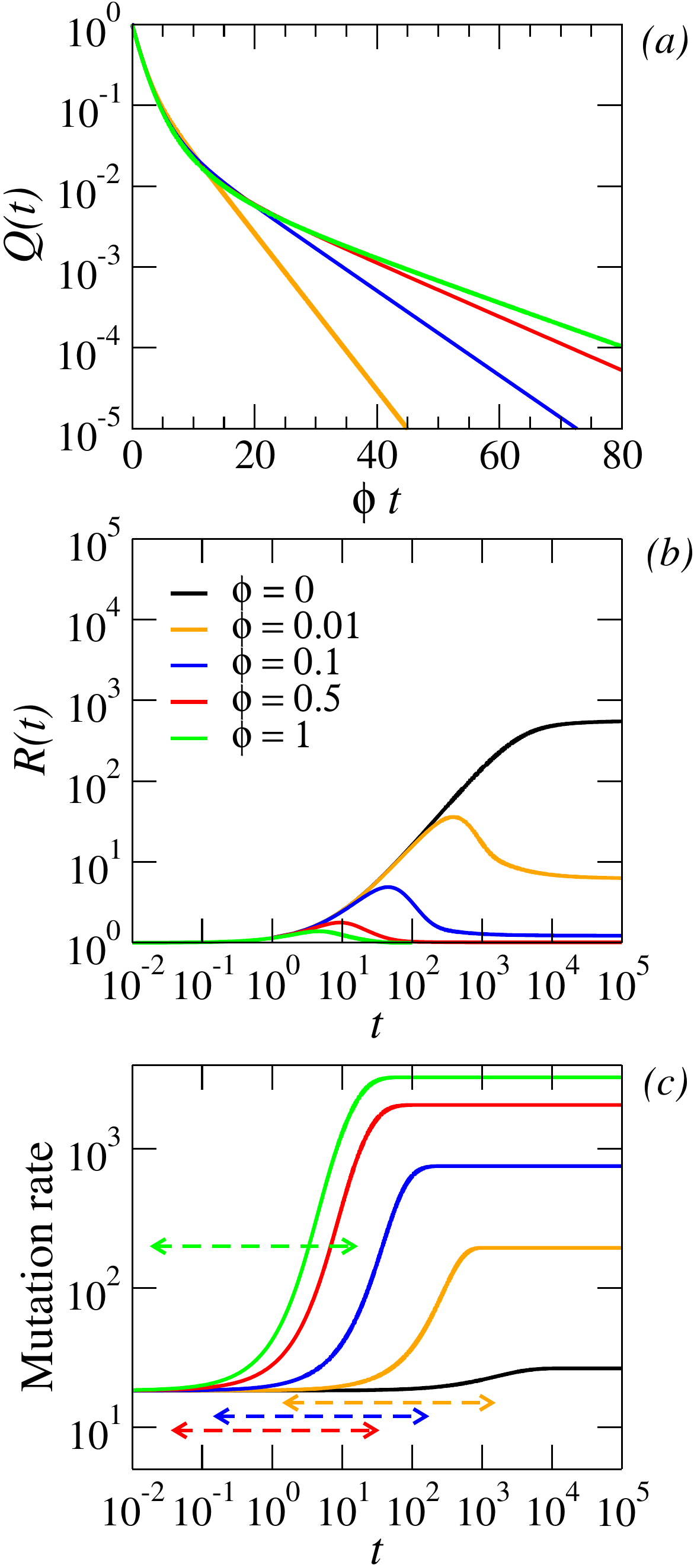}}
\caption{Dynamics of a RW on a network with a constant degree distribution,
with parameters described in the main text.
(a) Probability $Q(t)$ that the RW stays on the network after a time $t$. 
(b) Overdispersion of the network. (c) Mutation rate.  
 The assortativity is $r\approx 0.9997$. For the sake of comparison, these plots 
share scale with Fig.~\ref{fig:2degnetwork}.}
\label{fig:HDNnetwork}
\end{figure}

Figure~\ref{fig:HDNnetwork}(a) shows the presence of many different time scales
that become dominant at different times and cause a systematic bending of
$Q(t)$. The visibility of the different time scales is related to the presence
of significant fractions of nodes of different degrees together with the
canalization of the dynamics towards regions of increasingly higher
connectivity---a consequence of assortativity. This behaviour differs from
that of the two-degree network, where the curvature region was small and due to
the cross-over between the only two time scales involved. Overdispersion can
reach very high values, even once mutation-selection equilibrium has been
reached, and so does the mutation rate, see Figs.~\ref{fig:HDNnetwork}(b) and (c). 

In a network of this kind, the larger $\phi$, the less diverse is the set of
different degrees visited by the RW in the asymptotic state. This responds to the fact
that trajectories that survive for long times are preferentially concentrated in  
highly connected regions. As $\phi$ decreases, nodes of lower
degree become more competent at trapping the dynamics, causing two effects.
First, the asymptotic time scale $\alpha_{\rm min}$ changes, decreasing in absolute 
value with
$\phi$. Secondly, differences between the trajectories are less pronounced the
larger $\phi$, such that the maximum in the overdispersion and the largest 
acceleration of the clock are reached at earlier times. Third, the asymptotic values 
of $R(t)$ decrease with $\phi$, and those of the mutation rate increase.
At mutation-selection equilibrium, however, nodes with the highest connectivity
cannot fully trap the RW, such that nodes of lower degree are visited
frequently enough to yield $R(t) > 1$, attaining larger values the lower $\phi$.
In the asymptotic state, thus, the RW is more delocalized (in terms
of the visited degree) the lower the value of $\phi$.  
The fact that, at high $\phi$, those RW remaining on the NN preferentially visit 
nodes of high degree is responsible for the larger ticking rates observed---reaching 
values orders of magnitude bigger than initially.

\subsection{RNA networks}

Secondary structure RNA neutral networks (here RNA networks for short) are a
paradigm of the genotype-phenotype map, and many important properties on their
topology and on the GP mapping are known in detail. An RNA network is formed in
principle by all the sequences that yield the same minimum free energy
secondary structure. Usually, links are drawn between pairs of sequences in the
set that differ in a single point mutation. In general, this leads to a number
of disjoint connected components.  Since dynamics are well defined on connected
networks, we will assume that we work with subsets of the GP map that fulfill
this condition.  Among other properties known, there is a broad variability in
the number of sequences that fold into a given phenotype for a fixed sequence
length (several orders of magnitude)~\cite{schuster:1994,stich:2008}; the
average degree of an RNA network grows logaritmically with its size (number of
genotypes leading to the same phenotype) ~\cite{aguirre:2011}, and RNA networks
are highly assortative, with an index $r$ that increases on average with size
and approaches 1~\cite{aguirre:2011}. In this section we will discuss results
obtained for RNA networks of sequences of length $12$. The GP map for this case
is fully known, with networks of sizes spanning four and a half orders of
magnitude~\cite{aguirre:2011}.

\begin{figure}[ht]
\centerline{\includegraphics[width=70mm]{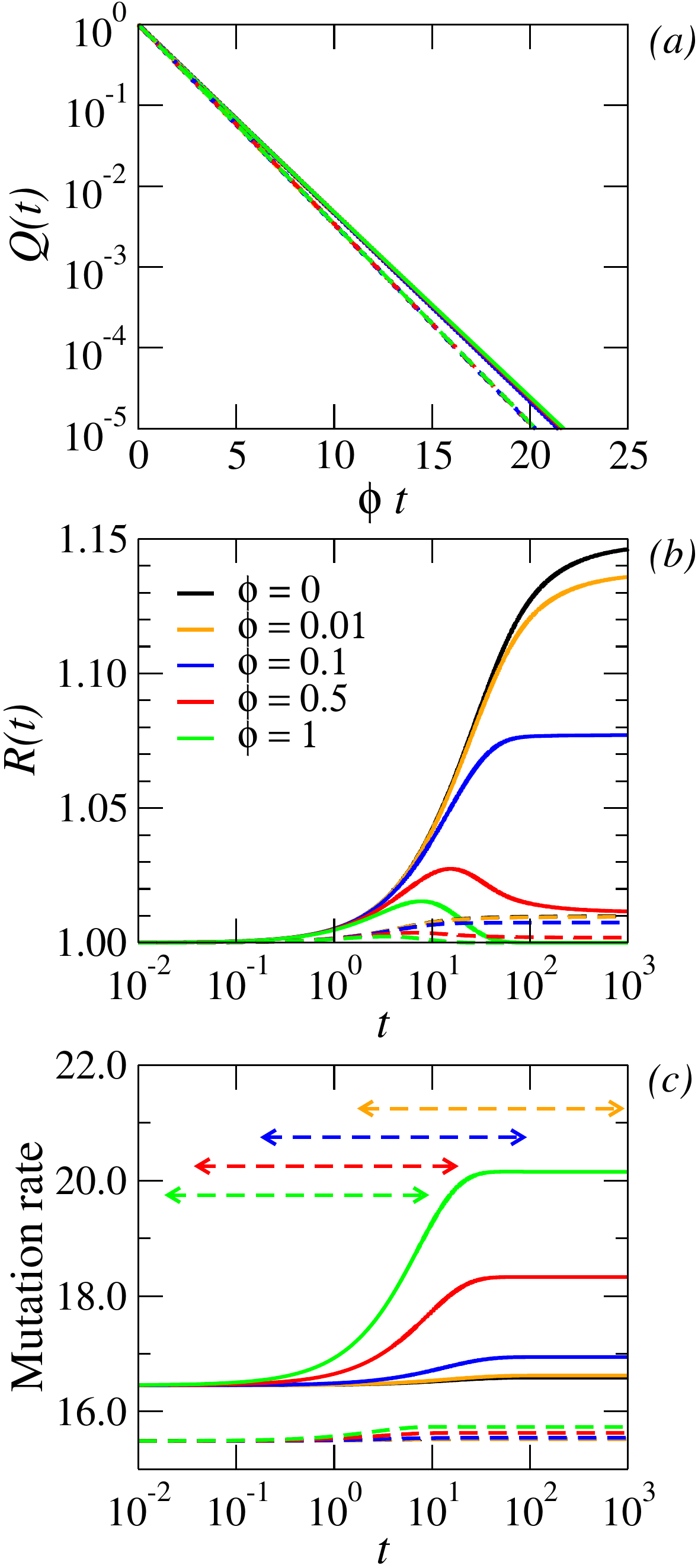}}
\caption{Dynamics of a RW on two different RNA secondary structure neutral
networks with $z=36$. Sizes are $M=1965$ (dashed lines) and $M=21908$ (solid
lines).  (a) Probability $Q(t)$ that the RW stays on the network after a time
$t$.  (b) Overdispersion of the network. (c) Mutation rate. Time 
$t_{1\%}(\phi)$ is only shown for the largest network. The assortativity is 
$r>0.90$. For the sake of comparison, these plots share scale with 
Fig.~\ref{fig:ERnetwork}.}
\label{fig:RNAnetwork}
\end{figure}

We have chosen two connected neutral networks formed by genotypes that belong
to different phenotypes. In dot-bracket notation, these are {\tt ...((.....))}
and {\tt (((.....))).}; the first phenotype can be obtained from 14,675
different sequences which are grouped into four different connected
subnetworks. As example we have chosen a connected component of size $M=1,965$.
The second phenotype is more abundant, corresponding to the minimum free energy
folded state of 142,302 sequences. As a second example we have chosen one of
its largest 17 connected subnetworks, of size $M=21,908$. Curves for $Q(t)$, $R(t)$, 
and the mutation rate are plotted in Fig.~\ref{fig:RNAnetwork}. As can be seen, 
the behaviour
of these networks is quantitatively comparable to the Erd\"os-Renyi random
network. In RNA, there is a monotonic change in the slower time scale and in
the equilibrium value of overdispersion with network size. This responds to the
particular relationship between size and degree exhibited by secondary
structure RNA neutral networks, where $\langle k \rangle \propto \ln M$. The
connection between the slower time scale and average degree can be made explicit 
in the calculations for random networks (Eqs.~\ref{eq:rhoER} and~\ref{eq:QtER}),
and despite the high assortativity of RNA networks, it also shows up here. The
degree distribution $p(k)$ of RNA neutral networks is very peaked for $M$
small, though its variance increases with $M$ faster than it does for a Poisson
degree distribution~\cite{aguirre:2011}.  Whether these differences lead, for
realistic values of $M$, to significant departures from the behaviour of random
networks remains to be seen.  

\section{Discussion}

Certain features of the dynamics of homogeneous populations on NN are 
generic, resulting from two ubiquitous topological properties: heterogeneity and 
assortativity. In general, the dynamical process that begins when the population 
enters the neutral network and finishes when mutation-selection equilibrium is 
achieved is characterized by as many different time scales as different degrees 
of robustness are found in the NN. These time scales successively show up along 
the transient and affect the process of mutation fixation. As time elapses, 
regions of increasingly higher degree are visited, causing the entrapment of the 
population. 
The actual time spent on a neutral network 
should increase as the difference between the fitness of the current phenotype and that 
of neighbouring alternatives grows, since fitter phenotypes trap populations for longer 
times.

Some of the features here described directly affect evolutionary dynamics on
realistic neutral networks and, therefore, quantitative properties of the
actual mutation substitution process. We have used as example RNA secondary
structure neutral networks since this is the only case we are aware of where
exhaustive studies of the genotype-phenotype map encompassing the whole space
of genotypes have been performed~\cite{gruner:1996,aguirre:2011}. Still, the
networks we may use to compute quantities such as overdispersion are limited by
current computational power. The average size of an RNA neutral network for
sequences of length $l$ grows as $0.673 l^{3/2} 2.1636^l$, which is an
astronomically large number even for moderate values of $l$~\cite{schuster:1994}. 
For instance, a natural tRNA 76nt long should have a neutral network formed by 
over $10^{28}$ different sequences. There is partial evidence that, as the length
of genotypes increases, the degree distribution of the corresponding neutral 
networks broadens, while assortativity grows with size~\cite{aguirre:2011}. If 
this is so, several different time scales should be visible and relevant in the 
dynamic process of mutation fixation in any realistic scenario. An additional 
example is provided by the neutral networks of natural proteins with known folds 
that have been computationally studied. Thought the techniques used only permit 
a partial exploration of those immense neutral networks, the degree 
distributions obtained are remarkably broad~\cite{bastolla:2003}. There is 
sufficient evidence as well that protein neutral networks are highly
assortative~\cite{bornberg-bauer:1999}. 

Network heterogeneity, together with stochasticity in the appearance and fixation of 
mutations, translates into potentially large values of $R(t)$. In most cases, 
overdispersion changes non-monotonically, presenting a maximum value at intermediate 
times, before equilibrium is reached. The value of evolutionary
time in our model yields an upper bound to the number of mutations accumulated
up to time $t$, since we have defined time in units of $\mu^{-1}$, where $\mu$
represents the rate of attempted mutations. Therefore, typical trajectories 
accumulate between a few and a few tens of mutations before leaving the network, 
a range where the strongest variation in $R(t)$ and mutation rates occurs. 
The non-monotonicity of $R(t)$ has been observed in realistic models of protein
evolution~\cite{bastolla:2003}. In an exhaustive analysis of thousands of protein 
sequences in several Drosophila species~\cite{bedford:2008}, it has been demonstrated 
that $R(t)$ is time- and gene-dependent, and that it should be described as a function 
of a phylogeny's level of divergence. The former observations are in full 
qualitative agreement with our results.

For any time $t$, $R(t)>1$ as long as genomes 
with different robustness (nodes with different degree) are visited with non-zero 
probability. There might be some extreme cases, as the two-degree networks we have 
analysed, when only nodes with the same degree are visible to the population at 
mutation-selection equilibrium. In those special situations (and, trivially, if the 
network is homogeneous) $R(t)=1$ is recovered, in agreement with Kimura's initial 
guess on the Poissonian nature of the substitution process. However, this is a very 
atypical situation that cannot be expected to occur in any realistic case. That network 
heterogeneity increases overdispersion had already been acknowledged 
before~\cite{raval:2007}. What had not been realized is the fact that this effect is 
stronger during transients; that transients are unavoidable, because evolution on NN 
begins preferentially in regions of low robustness; and that---depending on the value 
of $\phi$---equilibrium may be virtually unreachable before a mutation hits a 
fitter phenotype, driving the process again far from equilibrium.
Indeed, the number of mutations that have to be accumulated to attain equilibrium 
vary from slightly less than hundred to a few thousands, a situation not reached by
any typical trajectory unless $\phi$ is close to 1. It would be interesting to consider
other mutational mechanisms, such as recombination, which might affect the nature of 
equilibrium and the length of the transient. For example, it has been shown that, under
recombination, the mutation-selection equilibria of populations replicating on NN 
display a higher degree of robustness (are thus more entrapped) than those corresponding
to simple point mutations~\cite{szollosi:2008}.

There might be some natural systems where the accumulation of mutations,
and therefore the transient time before equilibrium is achieved, proceeds at a pace
significantly higher than that in most genes or proteins, namely RNA viruses, viroids, 
and miRNAs. These three systems are characterized by high mutation rates and, in RNA 
viruses and miRNAs, a correlation between selection for thermostability and increased 
robustness to mutations has been detected (a phenomenon called plastogenetic 
congruence)~\cite{domingo-calap:2010,szollosi:2009}. Viroids, which share several 
structural properties with miRNAs, also increase robustness along 
evolution~\cite{sanjuan:2006} and present a limited effect of mutations on their 
structure~\cite{manrubia:2013b}. The higher robustness of these systems with respect to
randomized counterparts is an indirect evidence of phenotypic entrapment and, therefore,
of populational states closer to mutation-selection equilibrium. Their corresponding
phylogenies might be excellent testing beds for the theoretical results here derived.

Evolutionary dynamics on heterogeneous and assortative networks with phenotype-dependent 
size has important consequences regarding the molecular clock~\cite{zuckerkandl:1965}. 
First, there is an increase in the rate at which mutations are fixed as time elapses, 
with an occasional maximum in its value during the transient. When the population first 
enters the network, the probability that it stays is relatively low.  However, if it
remains, the neutral mutation that is consequently fixed does it in a typical
time that depends on the robustness of the current genotype. Since robustness
systematically increases along the transient in assortative networks, the age of a 
population which has accumulated few neutral mutations will be systematically 
underestimated if a regularly ticking molecular clock is assumed. This 
prediction is in agreement with observations of a steady increase in $R(t)$ along 
phylogenies~\cite{bedford:2008} where, moreover, a gene-dependent $R(t)$ was identified.
Second, the ticking 
rate of the clock depends on $\phi$, that is, on the fitness value of the current 
phenotype (or NN) relative to neighbouring phenotypes: the smaller $\phi$, the higher 
the trapping power of the network, a situation that causes larger overdispersion
and lower asymptotic mutational rates.  
Finally, even sequences of the same length may belong to neutral networks that
differ in size in several orders of magnitude. The ticking rate of the clock is
unavoidably dependent on robustness, which in its turn is a function of
phenotype size in all genotype-phenotype maps 
studied~\cite{greenbury:2014,aguirre:2011,li:1996,bloom:2007,dallolio:2014}. 
Reliable estimations of neutral network sizes, at hand with appropriate computational
algorithms~\cite{jorg:2008}, should aid in the calibration of neutral evolutionary 
rates. This information could be combined with knowledge of $R(t)$ in
homologous genes with the same function (thus presumably characterized by the same NN).
Finally, the $R(t)$ and the NN sizes of different genes in diverging lineages could be 
compared to disentangle time-dependent changes in $R(t)$ from variation due to phenotype
sizes, extending studies such as those in~\cite{bedford:2008}.

The results here obtained do not contradict observations revealing a decrease in 
the ticking rate of the molecular clock along phylogenetic 
branches~\cite{pereira:2006,ho:2011}, since most fixed mutations are in those cases 
subject to selection. In this sense, our model addresses a partial aspect of 
phylogenies, namely strictly neutral evolution. A tractable model of how NN of
different fitness intermingle in the space of genomes may bring about the formulation 
of mean-field dynamical models, as the one here presented, that simultaneously take into 
account neutral drift and selection without discarding the complex architecture of the 
space of genotypes. Actually, it is plausible that the integration of several different 
structural features of phenotypes may clarify why the molecular clock depends so 
strongly on specific cases analysed \cite{kumar:2005}. Indeed, there are several 
mechanisms known to affect the uniformity of rates of molecular 
evolution~\cite{lio:1998}. In the light of the results here presented, also the 
particular topology of neutral networks, the degree of adaptation of populations, 
and the abundance of the current phenotype should be taken into account when 
calibrating molecular clocks. 

\section{Acknowledgements}

The authors acknowledge the financial support of the Spanish Ministerio de
Econom\'{\i}a y Competitividad under projects FIS2011-27569 and PRODIEVO
(FIS2011-22449), and of Comunidad de Madrid under project MODELICO-CM
(S2009/ESP-1691). They are indebted to Ricardo Azevedo and Ugo Bastolla for 
constructive criticism on the ms. 

\appendix

\section{Appendix --- Solving for $Q(t)$ in general}
\label{app:A}

According to eq.~\eqref{eq:Qgeneral}, to obtain $Q(t)$ we first need to compute
$\vP$. Except for very specific choices of $p_{\text{nn}}(k|j)$ and $\phi$, the
master equation \eqref{eq:mastermat} needs to be solved numerically. We will
prove here that matrix $\vF-\vA$ has $s$ (the cardinal of set $\mathcal{S}$)
different real eigenvalues---which justifies formula \eqref{eq:Qgeneral}.

Let us define $\vM(x)\equiv\vF-x\vA$, where $x$ is a complex variable. Standard
perturbation theory~\cite{kato:1980} tells us that the eigenvalues of $\vM(x)$
form a set of analytic functions of $x$ with constant cardinality, except for a
finite set of \emph{singular} points in the complex plane---where at least two
of the eigenvalues coalesce. But $\vM(0)=\diag\big(\phi+(1-\phi)k/z\big)$,
$k\in\mathcal{S}$, so it has $s$ different real eigevalues ($s$ being the
cardinal of $\mathcal{S}$), provided $\phi\ne 1$. Therefore $\vM(x)$ will also
have $s$ different eigenvalues---hence will be diagonalizable---except when $x$
is one of those singular points. Now, $x=1$ will not be a singular point for
almost any probability distributions $p(k|j)$.  But if $p(k|j)$ happens to be
one of those singular distributions, then the analysis to come is still valid
for $x=1+\epsilon$, for arbitrarily small $\epsilon$ (because the singular
points are isolated in the complex plane).  Thus we can eventually take the
limit $\epsilon\to 0$ without changing the conclusions.

We now show that all eigenvalues of $\vM(x)$ are real. To that aim let us
introduce the (non-singular) matrix $\vD=\diag\big((z/k)p(k)\big)$,
$k\in\mathcal{S}$. Then the elements of $\vT(x)\equiv\vM(x)\vD$ are $[(1-\phi)+
\phi z/k]p(k)\delta_{kj}-xp_{\text{nn}}(k,j)$, where $p_{\text{nn}}(k,j)$ is
the fractions of links of the NN joining two nodes of degrees $k$ and $j$.
Hence $\vT(x)$ is a symmetric matrix and therefore all its eigenvalues are
real.  But then so are the eigenvalues of $\vM(x)$ because
$\vM(x)=\vT(x)\vD^{-1}$, which has exactly the same eigenvalues as the
symmetric matrix $\vD^{-1/2}\vT(x)\vD^{-1/2}$.

Let $\vu_j$ and $\vv_j$, $j\in\mathcal{S}$, be the right and left eigenvectors
of $\vM(1)$ corresponding to the eigenvalue $\alpha_j$. They verify the
orthogonality condition $\vv_k\cdot\vu_j\propto\delta_{kj}$. Then, the solution
\begin{equation}
\vP(t)=\sum_{k\in\mathcal{S}}\frac{\vv_k\cdot\vP(0)}{\vv_k\cdot\vu_k}
e^{-\alpha_kt}\vu_k.
\end{equation}
Substituting this result into \eqref{eq:Qgeneral} yields \eqref{eq:Q}.

\section{Appendix --- $Q(t)$ and $R(t)$ for random networks}
\label{app:B}

To obtain $Q(t)$ we left-multiply \eqref{eq:mastermat} by $\vkappa^{\trsp}$ to
get
\begin{equation}
\frac{d}{dt}\vkappa\cdot\vP=-(1-\rho)\vkappa\cdot\vP, \qquad 
\rho\equiv\frac{\langle k^2\rangle}{z\langle k\rangle},
\label{eq:rhoER}
\end{equation}
whose solution is simply
\begin{equation}
\vkappa\cdot\vP(t)=\vkappa\cdot\vP(0)e^{-(1-\rho)t}, \qquad
\vkappa\cdot\vP(0)=\frac{(1-\rho)\langle k\rangle}{z-\langle k\rangle}.
\label{eq:kappaP}
\end{equation}
Substituting into \eqref{eq:Qgeneral} finally yields \eqref{eq:QtER}.

As for overdispersion, from eqs.~\eqref{eq:Gthth} we can derive
\begin{equation}
\begin{split}
\dot m_1 =& -m_1+G_1+\vkappa\cdot\vP, \\
\dot G_1 =& -(1-\rho)G_1+\rho\,\vkappa\cdot\vP, \\
\dot m_2 =& -m_2+G_2+2G_1+\vkappa\cdot\vP, \\
\dot G_2 =& -(1-\rho)G_2+\rho\,(2G_1+\vkappa\cdot\vP).
\end{split}
\label{eq:mG}
\end{equation}
with the initial condition $m_1(0)=m_2(0)=G_1(0)=G_2(0)=0$. Combining the
former two equations and the latter two we obtain
\begin{equation}
\frac{d}{dt}(G_i-\rho m_i) = -(G_i-\rho m_i), \quad i=1,2,
\end{equation}
which, given the null initial conditions, yield $G_i=\rho m_i$. Therefore
the four eqs.~\eqref{eq:mG} reduce to
\begin{equation}
\begin{split}
&\frac{d}{dt}m_1 = -(1-\rho)m_1+\vkappa\cdot\vP, \\
&\frac{d}{dt}(m_2-m_1) = -(1-\rho)(m_2-m_1)+2\rho m_1.
\end{split}
\end{equation}
Using \eqref{eq:kappaP}, this linear system can be readily solved:
\begin{equation}
\begin{split}
m_1(t) =& \frac{(1-\rho)\langle k\rangle}{z-\langle k\rangle} te^{-(1-\rho)t}, \\
m_2(t) =& \frac{(1-\rho)\langle k\rangle}{z-\langle k\rangle} (\rho t^2+t)
e^{-(1-\rho)t},
\end{split}
\end{equation}
from which overdispersion can be obtained as \eqref{eq:overdispersionrandom}.



\bibliographystyle{unsrtnat} \bibliography{JRSocInterface}


\section{Figure caption}

{\bf Short title:} Neutral networks and the molecular clock

\end{document}